\documentclass[11pt,a4paper]{article}
\usepackage{jheparxiv}
\usepackage[utf8x]{inputenc}
\usepackage{amsmath}
\usepackage{amsfonts}
\usepackage{amssymb}
\usepackage{latexsym}
\usepackage{mathrsfs}
\usepackage{braket}		
\usepackage{graphicx}
\usepackage{color}
\usepackage{xcolor}
\usepackage{slashed}
\usepackage{twistor}
\usepackage[all]{xy}
\usepackage{tikz-cd}
\usepackage{mathtools}
\usepackage{tikz-feynman}
\usepackage{bm}
\usepackage{color}
\usepackage{tikz}
\usepackage{braket}
\usepackage{float}
\usepackage[]{todonotes}

\makeatletter
\renewcommand{\todo}[2][]{%
    \@todo[caption={#2}, #1]{\begin{spacing}{0.5}#2\end{spacing}}%
} 
\makeatother 
\usepackage{setspace}

\title{The ultrarelativistic limit of Kerr}
\author[1]{Tim Adamo,}
\author[1]{Andrea Cristofoli,}
\author[2]{Piotr Tourkine}

\affiliation[1]{School of Mathematics and Maxwell Institute for Mathematical Sciences\\
University of Edinburgh, EH9 3FD, United Kingdom}

\affiliation[2]{LAPTh, CNRS et Universit\'e Savoie Mont-Blanc \\
9 Chemin de Bellevue, F-74941 Annecy, France}

\emailAdd{t.adamo@ed.ac.uk}
\emailAdd{acristof@exseed.ed.ac.uk}
\emailAdd{tourkine@lapth.cnrs.fr}

\abstract{The massless (or ultrarelativistic) limit of a Schwarzschild black hole with fixed energy was determined long ago in the form of the Aichelburg-Sexl shockwave, but the status of the same limit for a Kerr black hole is less clear. In this paper, we explore the ultrarelativistic limit of Kerr in the class of Kerr-Schild impulsive pp-waves by exploiting a relation between the metric profile and the eikonal phase associated with scattering between a scalar and the source of the metric. This gives a map between candidate metrics and tree-level, 4-point scattering amplitudes. At large distances from the source, we find that all candidates for the massless limit of Kerr in this class do not have spin effects. This includes the metric corresponding to the massless limit of the amplitude for gravitational scattering between a scalar and a massive particle of infinite spin. One metric, discovered by Balasin and Nachbagauer, does have spin and finite size effects at short distances, leading to a remarkably compact scattering amplitude with many interesting properties. We also discuss the classical single copy of the ultrarelativistic limit of Kerr in electromagnetism.}

\begin{document}

\clearpage
\maketitle

\section{Introduction}

The use of scattering amplitudes to derive classical solutions to the Einstein field equations has a long history. The first attempt in this direction goes back to work by Duff~\cite{Duff:1973zz}, where the Schwarzschild solution in harmonic coordinates was derived to order $G^2_{N}$. This approach can also be implemented through unitarity techniques~\cite{Neill:2013wsa}, and similar methods can be used to obtain other solutions such as Reissner-Nordstr\"om and Kerr~\cite{Sardelis:1973em} as well as quantum corrections, higher-order corrections and higher-dimensional generalizations~\cite{Bjerrum-Bohr:2002fji,Mougiakakos:2020laz,Jakobsen:2020ksu}. Other approaches have focused instead on effective field theory techniques for the derivation of these solutions~\cite{Goldberger:2004jt,Goldberger:2016iau} or the use of one-point functions for the extraction of the large distance behaviour~\cite{KoemansCollado:2018hss}. More recently, the relevance of three-point amplitudes for the derivation of classical solutions has been emphasised within the framework of classical observables~\cite{Kosower:2018adc,Cristofoli:2021vyo}, using off-shell techniques in Lorentzian signature~\cite{Cristofoli} and on-shell ones in split signature~\cite{Monteiro:2020plf,Monteiro:2021ztt}, while twistor methods have also been employed to provide interesting results related to classical solutions \cite{Justin-notes,Guevara:2021yud}.

Despite the many insights that these links have provided, so far they have not shed any new light on the space of solutions to the Einstein equations. This is not terribly surprising; for instance, all algebraically special type D, stationary, axisymmetric solutions -- precisely the sort generated by most of the amplitudes-based techniques mentioned above -- were classified long ago~\cite{Plebanski:1976gy}. In this paper, we will use on-shell, amplitudes-based methods to investigate a question in classical general relativity where the answer does \emph{not} appear to be widely known or agreed upon in the literature, namely: what is the massless limit of the Kerr metric with fixed, finite energy?

\medskip

In the case of the Schwarzschild metric, the answer to this question is well-known and given by the Aichelburg-Sexl shockwave metric~\cite{Aichelburg:1970dh}, sourced by an ultraboosted scalar particle~\cite{Dray:1984ha}. By contrast, taking the massless limit of the Kerr metric is more ambiguous. In the first case, the Kerr metric is characterised by two parameters: a mass $M$ and the spin parameter $a$, which are constrained by the extremality bound $M\geq a$. So naively, any interesting massless limit (i.e., with spin effects) with $M=0$, $a\neq0$ will violate the extremality bound\footnote{This is true even in the limit where the energy is also taken to zero; contrary to na\"ive expectations, this results in a wormhole solution which is only locally flat, rather than Minkowski spacetime~\cite{Gibbons:2017djb}.}. Nevertheless, the massless limit -- or ultraboost -- of a finite mass metric is sufficiently slippery that there seems to be no consensus as to whether or not spin effects can persist in the massless limit of Kerr, and if so whether they extend to large distances.

We will attempt to clarify this situation by focusing on a particular class of metrics: Kerr-Schild impulsive pp-waves. In particular, we wish to ascertain: do candidates for the ultrarelativistic limit of Kerr in this class exhibit spin effects, and if so do they extend to long distances (i.e., far from the source of the metric)? Surprisingly, these questions can be answered using the on-shell data of scattering amplitudes. We show that there is a one-to-one correspondence between metrics in this class and an eikonal phase associated to gravitational scattering between a probe scalar and the source of the metric in quantum field theory (QFT).

We use this correspondence to consider all ultraboosts of the Kerr metric in the Kerr-Schild impulsive pp-wave class, as well as other candidates in the literature. We find that ultraboosts of the line element (both parallel and perpendicular to the axis of rotation)~\cite{Balasin:1995tj,Podolsky:1998rp,Barrabes:2003up} are encoded by the massless limit of the scattering amplitude between a scalar and an infinite spin particle~\cite{Guevara:2018wpp,Guevara:2019fsj}. In this case, we find that the resulting metric is actually equivalent to the shockwave, meaning that there are no spin effects at all. We then analyze other candidates in the literature: one by Ferrari and Pendenza obtained by another ultraboosting procedure of the metric~\cite{Ferrari:1990tzs}, and another by Balasin and Nachbagauer obtained by ultraboosting the source of Kerr parallel to the axis of symmetry~\cite{Balasin:1994tb}. We show that the source of the former is not compactly supported, ruling it out as a viable candidate.

The Balasin-Nachbagauer solution, on the other hand, does have a compactly supported source with finite size. At large distances, the spin effects switch off, but they are present at short scales. This induces a four-point scattering amplitude between a scalar and a compact disc of null dust which takes a very simple form:

\be\label{BNamp*}
A_{4}=\frac{\kappa^2\,s^2}{8\,t}\, \left(\frac{\sin(a \sqrt{-t})}{a\,\sqrt{-t} }+\cos(a \sqrt{-t}) \right)\,,
\ee
where $\kappa=\sqrt{32 \pi G_{N}}$ is the gravitational coupling constant, $s$ is the Mandelstam variable related to the energy in the center of mass frame, $t$ is the square of the exchanged momentum and $a$ is the spin parameter. As we shall see, in contrast to the gravitational scattering of point particles, the finite size effects in this amplitude lead to non-trivial scales for the onset of spin effects in the tree-level and eikonal regimes, as well as regularized high-energy behaviour.

\medskip

The paper is organized as follows: in Section~\ref{Sec:KS-Amp} we derive the correspondence between Kerr-Schild impulsive pp-wave metrics and scattering amplitudes via an eikonal phase. Section~\ref{Sec:Metrics} applies this correspondence to candidates for the massless limit of Kerr which can be derived from the scattering amplitude between a scalar and a particle with infinite spin~\cite{Guevara:2018wpp,Arkani-Hamed:2019ymq,Guevara:2019fsj} or are otherwise in the literature. Section~\ref{Sec:BN} explores the structure of the scattering amplitude associated to the Balasin-Nachbagauer metric, which we argue is the most interesting massless limit of Kerr in the class of metrics we study. Finally, we consider the classical single copy of the ultrarelativistic limit of Kerr in Section~\ref{Sec:SingleCopy}. Throughout the paper, we use the mostly-minus signature convention for the metric and natural units with $\hbar=1$.


\section{Impulsive pp-waves from on-shell data}
\label{Sec:KS-Amp}

In this paper, we wish to look at candidates for the massless limit of the Kerr metric in the class of \emph{Kerr-Schild, impulsive pp-waves} (cf., \cite{Ehlers:1962zz,Stephani:2003tm,Griffiths:2009dfa,Blau:2011}); these are Kerr-Schild metrics with a covariantly constant null Killing vector $n^{\mu}$, where the non-trivial metric components are localized on the $n^\mu$ lightfront. In lightfront coordinates $x^{\mu}=(x^{-}, x^{+}, x^{\perp})$, we can take $n^{\mu}=\delta^{\mu}_{+}$ and metrics in this class can be put into the form
\be\label{ppwave}
\d s^{2}=2\,\d x^{-}\,\d x^{+}-(\d x^{\perp})^2+ \delta(x^-)\,f(x^{\perp})\,(\d x^{-})^2\,,
\ee
where the profile function, $f(x^{\perp})$, determines the curvature tensor and hence the source of the metric. By considering the scattering of a probe scalar in spacetimes of this class, one obtains a direct relationship between the profile $f(x^{\perp})$ and the scattering amplitude between the scalar and the source of the metric in the form of an eikonal phase.

In this section, we describe how this relationship emerges in general, and illustrate it with the concrete example of the Aichelburg-Sexl shockwave, where the amplitude side of the relation corresponds to scattering between the probe scalar and a massless scalar which is the source of the shockwave.


\subsection{The profile/eikonal phase relation}

Consider the ultra-relativistic limit of a four-dimensional black hole solution, where $n^{\mu}$ is the momentum of the ultraboosted source (so $n^2=0$). The most general line element in Kerr-Schild coordinates that arises from this ultraboost is~\cite{Shore}
\begin{equation}\label{eq:gen-line-element}
\d s^{2}=\eta_{\mu\nu}\,\d x^{\mu}\,\d x^{\nu}+\,\delta(n \cdot x)\, f(x^{\perp})\, n_{\mu}\, n_{\nu}\, \d x^{\mu}\, \d x^{\nu} \, .
\end{equation}
Let $\bar{n}^{\mu}$ be the null vector such that $\eta_{\mu\nu} n^{\mu}\,\bar{n}^{\nu}=1$; the pair $\{n^{\mu},\bar{n}^{\mu}\}$ defines two null directions in the spacetime, with the remaining two (independent) spacetime coordinates denoted by $x^{\perp}$. The profile function $f$ is a function only of the $x^{\perp}$ directions. Without loss of generality, we can choose the momentum of the ultraboosted source $n^{\mu}=\delta^{\mu}_{+}$ for $x^{\pm}:=\frac{1}{\sqrt{2}}(x^{0}\pm x^{3})$, in which case \eqref{eq:gen-line-element} becomes \eqref{ppwave}.

The metric \eqref{ppwave} has a single non-vanishing Ricci curvature component, and consequently a single stress tensor component
\be\label{Ricci}
R_{--}=8\,\pi\, G_{N}\: T_{--}=\frac{\delta(x^-)}{2}\,\partial^{2}_{\perp} f(x^{\perp})\,.
\ee
We say that a metric \eqref{ppwave} is an `\emph{admissible}' ultraboost of a black hole if the stress tensor $T_{--}$ is non-zero but compactly supported in the transverse variables $x^\perp$. This is a fairly minimal criteria, ruling out only those metrics of the form \eqref{ppwave} which are globally vacuum or have diffuse, null-dust-like sources -- neither of which is appropriate for the ultraboost of a black hole which started life as a vacuum solution to the Einstein equations outside of a spatially compact source.

\medskip

Since the curvature \eqref{Ricci} of these metrics is localized on the $x^-=0$ lightfront, there are well-defined in ($x^-<0$) and out ($x^->0$) regions of the spacetime, so it makes sense to consider classical scattering between these regions. It is a remarkable fact that the profile function $f(x^\perp)$ is uniquely defined by the scattering of a test scalar particle in the spacetime. To see this, consider the perturbative (in $G_N$) calculation of the impulse imparted to a massive test particle with initial momentum $p^{\mu}$ following a geodesic in \eqref{ppwave}. At leading order, this is given by
\begin{equation}
\begin{split}
\Delta p^{\mu}&=\frac{1}{2} \int_{\mathbb{R}} \d \sigma\, \partial^{\mu} h_{\alpha \beta}(x(\sigma))\, p^{\alpha} p^{\beta} \\
 & = \frac{p_+^2}{2}\int_{\R}\d\sigma\,\partial^{\mu}\left[\delta(x^{-}(\sigma))\,f(x^{\perp})\right]\,,
\end{split}
\end{equation}
where in the first line
\be\label{linmet}
h_{\alpha\beta}(x):=\delta(x^-)\,f(x^{\perp})\,n_{\alpha}\,n_{\beta}\,,
\ee
and the motion of the test particle is parametrized as
\begin{equation}
x^{\mu}(\sigma)=p^{\mu}\, \sigma+b^{\mu} \,, \qquad  b \cdot p=0 \,, \quad b \cdot n=0 \,.
\end{equation}
With this parametrization, the impulse has contributions from two terms
\be\label{impulse}
\Delta p^{\mu}=\left.\frac{p_+}{2}\, \partial^{\mu}  f(x^{\perp}(\sigma))\right|_{\sigma=0}+\frac{p_+^2}{2}\int_{\mathbb{R}} \d \sigma \, n^{\mu}\, \delta'(x^{-}(\sigma))\, f(x^{\perp}(\sigma)) \,,
\ee
corresponding to whether the derivative acts on the profile or the delta function.

Now, the scattering angle $\theta$ for the probe particle is given at leading order in the coupling by 
\begin{equation}\label{sa1}
\theta = \frac{\Delta p \cdot b}{p_{+} \,\mathbf{b}}+\cdots\,,
\end{equation}
where $p_+=p\cdot n$ and $\mathbf{b}$ the modulus of the impact parameter along its $\perp$-components. Since
\begin{equation}
\int_{\mathbb{R}} \d \sigma \, b \cdot n\, \delta'(x^{-}(\sigma))\, f(x^{\perp}(\sigma))=0\,,
\end{equation}
it follows that only the first term in \eqref{impulse} actually contributes to the scattering angle:
\begin{equation}
\Delta p \cdot b=\left.\frac{p_+}{2}\, b\cdot\partial  f(x^{\perp}(\sigma))\right|_{\sigma=0}\, .
\end{equation}
Using the fact that $\mathbf{b} \d\mathbf{b}=b_{\mu}\d b^{\mu}$, one obtains
\begin{equation}\label{sa2}
\theta =\frac{1}{2}\,\left.\frac{\d f(x^{\perp})}{\d \mathbf{b}}\right|_{x^{\mu}=b^{\mu}}+\cdots\,,
\end{equation}
for the scattering angle at leading order.

The same observable can be computed at leading order in the coupling and large impact parameter using eikonal methods~\cite{KoemansCollado:2019ggb,Cristofoli:2021jas}
\begin{equation}\label{sa3}
\theta =-\frac{1}{p_{+}}\frac{\d\chi_{1}(b^{\perp})}{\d\mathbf{b}}+\cdots\,,
\end{equation}
where $\chi_{1}(b^\perp)$ is the leading eikonal phase associated to scattering between a massive probe and a massless particle (with or without spin) corresponding to the source of the pp-wave metric. Equating the two formulae \eqref{sa2} and \eqref{sa3} leads to the following relation between the metric profile and the eikonal phase: 
\begin{equation}\label{eq:eik-pro-map}
f(x^{\perp})=-\frac{2\chi_{1}(x^{\perp})}{p_+} \, .
\end{equation}
Of course, at next-to-leading order in the coupling, the relation between the metric profile and eikonal phase should become non-linear (indeed, we will see hints of this later), but for the moment we need only consider the tree-level relationship \eqref{eq:eik-pro-map}.

Recall that the leading eikonal phase $\chi_1$ for any scattering process between two particles of masses $m$ and $M$ with momenta $p$ and $P$ is simply the inverse Fourier transform of the tree-level 4-point scattering amplitude (or Born amplitude) $A_{4}(q_\perp)$ (cf., \cite{Abarbanel:1969ek,Levy:1969cr,Wallace:1977ae}):
\be\label{eikphase}
\chi_{1}(x^{\perp})=\int\frac{\d^{4}q}{(2\pi)^2}\,\delta(2\,p\cdot q)\,\delta(2\,P\cdot q)\,\e^{\im\,q\cdot x}\,A_{4}(q)\,,
\ee
which is easily inverted
\be\label{4ptamp}
A_{4}(q_{\perp})=4\sqrt{(p \cdot P)^2-m^2 M^2}\int\d^{2}x^{\perp}\,\e^{-\im\,q_{\perp}x^{\perp}}\,\chi_{1}(x^{\perp})\,,
\ee
to give the tree-level amplitude from the eikonal phase.

\medskip

The main idea is now simple: \eqref{eq:eik-pro-map} gives a two-way map linking metric profiles $f(x^{\perp})$ with 4-point tree-level scattering amplitudes $A_{4}(q_{\perp})$, which for large impact parameter (or small momentum transfer) should correspond to the Born approximation of $2\to2$ tree-level scattering between the probe scalar and a null particle representing the source of the metric. This means that given any admissible metric of the form \eqref{ppwave}, we can read off an associated tree-level scattering amplitude. If this amplitude exhibits the behaviour expected for scattering between a scalar and a massless particle with the appropriate quantum numbers, then this is strong evidence that the given metric is a good candidate for the description of the associated ultraboosted black hole. Similarly, given a Born amplitude which is expected to give a tree-level QFT description of a scalar scattering off an ultraboosted black hole, then one can read off the associated metric profile to check that it is admissible.

In the case of the massless limit of Kerr, this gives a litmus test which any candidate metric of the form \eqref{ppwave} must pass, and also allows us to determine whether or not there are any spin effects. Similarly, by taking the massless limit of the four-point amplitude for scattering between a scalar and an infinite spin particle~\cite{Guevara:2018wpp,Guevara:2019fsj}, we are able to read off associated metric profiles and study their properties. Before embarking on this, it is illustrative to demonstrate how the relation \eqref{eq:eik-pro-map} works in practice with a concrete, unambiguous example: the massless limit of the Schwarzschild metric.


\subsection{An example: the massless limit of Schwarzschild}

Using the two-way relation between the metric profile and eikonal phase \eqref{eq:eik-pro-map}, we expect the massless limit of the Schwarzschild metric to correspond to a four-point scattering amplitude between a massive probe and a massless scalar with lightfront energy $P_-$. The tree-level amplitude for this process is given by
\be\label{Sch-tree}
A_{4}(q_\perp)=32 \pi G_{N} \frac{(p_+\,P_-)^2}{q_\perp^2}\,.
\ee
From this, the leading eikonal phase for the scattering process in position space is~\cite{tHooft:1987vrq,Kabat:1992tb,KoemansCollado:2019ggb,Adamo:2021rfq,Adamo:2022rmp} 
\begin{equation}\label{Sch-eik}
\chi_{1}(x^{\perp})=-4 G_{N}\, p_{+}\,P_{-}\,\log(\mu\,r) \, ,
\end{equation}
where $\mu$ is an arbitrary scale introduced to regularize IR divergences in the Fourier transform \eqref{4ptamp} and $r:=\sqrt{(x^{\perp})^2}$.

Feeding this into our fundamental relation \eqref{eq:eik-pro-map} gives the profile function
\begin{equation}\label{ASprofile}
f(x^{\perp})=8\, G_{N}P_{-}\,\log(\mu\,r)\,,
\end{equation}
which we immediately recognize as the Aichelburg-Sexl shockwave metric~\cite{Aichelburg:1970dh}. Of course, this is no surprise: the stress tensor for the shockwave is that of an ultraboosted scalar particle located at the origin in the transverse plane~\cite{Dray:1984ha}:
\be\label{ASsource}
T_{--}=P_{-}\,\delta(x^-)\,\delta^{2}(x^{\perp})\,,
\ee
which is precisely what one expects for the massless limit of the Schwarzschild metric.

It is worth emphasizing that this derivation does not involve any off-shell three-point functions as in~\cite{Cristofoli}, but only the knowledge of an on-shell eikonal phase in the probe limit. It is a more straightforward derivation, with the advantage of removing ambiguities coming from the use of off-shell amplitudes. 


\section{The ultrarelativistic limit of Kerr}
\label{Sec:Metrics}

Unlike the massless limit of the Schwarzschild metric -- which can be obtained fairly straightforwardly by ultraboosting the line element itself, ultraboosting the source, or from scattering amplitudes -- the massless limit of the Kerr metric is more ambiguous. To begin with, any limit which takes the mass $M\to0$ while leaving the spin parameter $a$ finite will violate the Kerr extremality bound and, at least naively, result in a naked singularity. But the null nature of the resulting solution could soften this singularity\footnote{For example, the curvature tensor of the shockwave has a delta-function singularity on the $x^-=0$ lightfront, so the metric is only Lipschitz-regular (in appropriate coordinates). Nevertheless, this singularity is fairly mild: the Kretschmann scalar vanishes everywhere and geodesics are continuously differentiable across the lightfront~\cite{Penrose:1972xrn,Ferrari:1988cc,Balasin:1996mq,Steinbauer:1997dw,Podolsky:1998my,Steinbauer:2006qi,Samann:2012xm,Lecke:2013lja}}, or one could simply accept the presence of a naked singularity, viewing the physical region of the resulting metric as being only at large distances. The question then remains: does the resulting metric have spin effects at large distances? 

In this section, we explore this question for various notions of the massless limit of the Kerr metric which can be defined using scattering amplitudes or have otherwise appeared in the literature. In all cases that are admissible, we find that there are no spin effects at large distances, where the geometry is identical to that of a shockwave.


\subsection{The massless limit of the spinning amplitude}
To begin, we consider the tree-level scattering amplitude between a scalar of mass $m$ and initial (final) momentum $p^{\mu}$ $(p^{\mu}\:')$ and a particle with infinite spin of mass $M$ and initial (final) momentum $P^\mu$ $(P^{\mu}\:')$. The Pauli-Lubanski vector for the spinning particle is
\begin{equation}
\label{spin-vec}
s_{\lambda}=\frac{1}{2\, M}\, \epsilon_{\lambda \mu \nu \rho} S^{\mu \nu}\, P^{\rho} \, ,
\end{equation} 
where $S^{\mu \nu}$ is the spin tensor. The leading classical limit of the tree-level interaction between the two particles - linear in $G_{N}$ and to all order in spin - is given by~\cite{Guevara:2018wpp,Guevara:2019fsj}:
\begin{equation}\label{eq:GOV-v2}
A_{4}(q)=8\pi\, G_{N}\, \frac{(p\cdot P)^2}{q^{2}} \sum_{\pm}(1 \pm v)^{2}\,\exp\left(\pm\im\, \frac{q \cdot [w * s]}{M}\right) \, ,
\end{equation}
where $q_{\mu}:=p_{\mu}-p'_{\mu}$ is the exchanged momentum and
\begin{equation}
w^{\mu \nu}:=\frac{2\, p^{[\mu} P^{\nu]}}{m\, M\,v\, \gamma(v)}\,, \qquad [w * s]_{\mu}:=\frac{1}{2}\, \epsilon_{\mu \nu \alpha \beta}\, w^{\alpha \beta}\, s^{\nu} \,.
\end{equation}
The relative velocity $v$ is defined through the gamma factor as
\begin{equation}
\gamma=\frac{1}{\sqrt{1-v^2}}=u\cdot U  \,,
\end{equation}
where $p^{\mu}=m u^{\mu}$ and $P^{\mu}=MU^{\mu}$.

There is now a significant amount of evidence that this amplitude describes the scattering between a probe scalar and the Kerr metric (at sufficiently large distances) to linear order in $G_N$ and all orders in spin~\cite{Chung:2018kqs,Maybee:2019jus,Arkani-Hamed:2019ymq,Bautista:2021wfy,Adamo:2021rfq}. Introducing the covariant impact parameter $b^{\mu}$, the eikonal phase associated with this amplitude becomes
\begin{equation}\label{GOVmass}
\chi_{1}(b)=-G_{N}\, \frac{p\cdot P}{v}\,\sum_{\pm}(1 \pm v)^{2}\, \log \sqrt{-\left(b \mp [w * a]\right)^{2}} \, ,
\end{equation}
where $a^{\mu}:=s^{\mu}/M$ is the re-scaled Pauli-Lubanski vector.

\medskip

Now, we wish to consider the massless limit with respect to the spinning particle, leaving an eikonal phase that can be fed into \eqref{eq:eik-pro-map}. Naively, this corresponds to the $v\to1$, $M\to0$ limit, while keeping the energy of the spinning particle fixed. However, the spin vector \eqref{spin-vec} also contains a factor of $M^{-1}$, requiring careful treatment in order to obtain a finite massless limit. This can be accomplished by demanding that the ring radius $a^{\mu}$ remains constant; while this may seem un-natural (indeed, it means the spin tensor vanishes in the massless limit), it can be interpreted as saying that the correct spin information is encoded in the massless limit in the finite vector $a^{\mu}$. Something similar is done in gluing constructions of spinning null metrics which keep a finite ring radius with vanishing angular momentum~\cite{Frolov:2022rbv}. 

In the massless limit where $P^{\mu}\to P_{-}\,n^{\mu}$, this leads to a finite eikonal phase: 
\begin{equation}
\lim_{M  \rightarrow 0}\chi_{1}(b_{\perp})=-4\,G_{N}\,p_{+}\,P_{-}\,  \log\left( \mu\,\sqrt{\left(b - \tilde{a}\right)_{\perp}^{2}}\right) \, ,
\end{equation}
where $\mu$ is an arbitrary mass scale, $\perp$ is the orthogonal plane to $p$ and $P$ and we define:
\be\label{tildea}
\tilde{a}^{\mu}:=P_{-}\,\epsilon^{\mu}{}_{\nu\rho\sigma}\,a^{\nu}\,u^{\rho}\,n^{\sigma}\,,
\ee
by keeping $a$ fixed in the massless limit. This can now be fed directly into \eqref{eq:eik-pro-map} to obtain a profile for the associated metric\footnote{Note that this simple expression could have been derived directly by using (3.25) in \cite{Guevara:2018wpp}, rather than (\ref{eq:GOV-v2}).}
\be\label{eq:GOV-gyraton}
f(x^{\perp})=8\,G_{N} P_{-}\,\log\left(\mu\,\sqrt{(x-\tilde{a})_{\perp}^2}\right)\,.
\ee
Note that by definition \eqref{tildea}, $\tilde{a}^{\mu}$ only has non-trivial components in the $\perp$-directions.

At this point, it is easy to study various properties of the metric defined by this profile, depending on the direction of the spin $a^{\mu}$ relative to the direction of the ultraboost. Remarkably, we will see that this corresponds with results for the direct ultraboost of the Kerr metric in the literature.


\subsection{Ultraboosts in various directions}

The residual spin vector $a^\mu$ in \eqref{eq:GOV-gyraton} is arbitrary, and can point in any direction. Thus, there are two cases to consider: the ultraboost is parallel or orthogonal to the direction of spin. The parallel case is very simple; in this case $\tilde{a}^{\mu}=0$ and the profile function immediately reduces to 
\begin{equation}\label{parallel}
f_{\mathbf{n}||\mathbf{a}}(x^{\perp})=  8\,G_{N}P_{-}\,\log (\mu\,r)\,,
\end{equation}
which we recognize as the (non-spinning) Aichelburg-Sexl shockwave \eqref{ASprofile}. In other words, in the case of a parallel ultraboost, we obtain a metric with no spin effects at all. This matches results in the literature for the ultraboost of the Kerr metric along the axis of rotation~\cite{Podolsky:1998rp,Barrabes:2003up} using various techniques.

In the case of an orthogonal ultraboost, the situation is slightly more subtle. Without loss of generality, we can assume that the spatial part of the spin vector $a^{\mu}$ points in the $x^1$-direction of the transverse plane: $\vec{\mathbf{a}}=a\,\vec{e}_{1}$. In this case, the profile function becomes
\begin{equation}\label{eq:GOV-perp}
f_{\mathbf{n} \perp \mathbf{a}}(x^{\perp})= 8\,G_{N}P_{-}\, \log\!\left(\mu\,\left|(x^1+a)^2+(x^2)^2\right|\right)\,.
\end{equation}
As in the parallel case, this also agrees with results in the literature for the ultraboost of Kerr orthogonal to the axis of rotation~\cite{Balasin:1995tj,Barrabes:2003up,Frolov:2022rbv}.

At first glance, it might seem that the resulting metric contains spin effects, as the profile function depends on $a$. However, this dependence on $a$ is easily removed by a linear diffeomorphism $x^1\to x^1-a$, after which the profile reduces to that of the non-spinning shockwave. Equivalently, the stress tensor associated to \eqref{eq:GOV-perp} is (before performing any diffeomorphism):
\be\label{PerpST}
T_{--}=P_{-}\,\delta(x^-)\,\delta(x^1+a)\,\delta(x^2)\,,
\ee
which is just the stress tensor for a massless particle of lightfront energy $P_-$, travelling in the $n^\mu$ direction and located at $x^{\perp}=(-a, 0)$ in the transverse plane. 

One can speculate that the origin for this shift of the source from its original center at $x^\perp=0$ lies in the ``unphysical'' ultraboost, for which the spin parameter $a$ was kept constant as the mass of the black-hole vanished, and the energy density is localised at the extremal point of rotation of the black-hole, where the rotation is parallel to the boost. This shift, as we discuss momentarily, results in an ambiguity in the definition of the impact parameter and related scattering angle. The transverse plane is just an affine space, with no physical meaning for the origin $x^{\perp}=(0,0)$, so we can simply shift the location of the origin to remove the dependence on $a$, which is precisely the diffeomorphism $x^1\to x^1-a$.

\medskip

So in both cases, we obtain a metric which is simply the Aichelburgh-Sexl shockwave. As this metric is sourced by a scalar (non-spinning) particle moving at the speed on light, 
anything resembling spin effects can only arise from an ambiguity in comparing the coordinate systems before and after the ultraboost. For instance, there are some examples in the literature where the orthogonal ultraboost is treated as if it has spin effects, despite being diffeomorphism-equivalent to the non-spinning shockwave. This seems to be centered around the computation of the scattering angle for the probe particle, and it is illustrative to explore this a bit further.


Using the formula \eqref{sa3}, the scattering angle at leading order is found to be~\cite{Barrabes:2003ey}
\be\label{GOVsa}
\theta=-\frac{4\,G_{N}\,P_{-}}{\mathbf{b}+a}\,,
\ee
where $\mathbf{b}=\sqrt{(b^{\perp})^2}$. Now, the leading-order formula for the scattering angle in the Kerr metric is~\cite{Boyer:1966qh}
\be\label{Kerrsa}
\theta^{\mathrm{Kerr}}=-\frac{4\,G_{N}\,M}{b+a}\,,
\ee
where $M$ is the Kerr mass and $b$ is the impact parameter. Thus, by identifying $\mathbf{b}$ with the physical impact parameter, it seems that $a$ serves to encode spin effects in \eqref{GOVsa} just as it does in Kerr.

However, the physical impact parameter encodes the distance to the source at closest approach, and since the source is located at $x^{\perp}=(-a,0)$, the physical impact parameter is actually $b=\mathbf{b}+a$. This leaves:
\be\label{GOVsa1}
\theta=-\frac{4\,G_{N}\,P_{-}}{b}\,,
\ee
which is nothing but the scattering angle in a shockwave and has no spin effects. Once again, the rescaling needed to obtain the physical result corresponds to the diffeomorphism which makes the equivalence with the shockwave manifest. One could worry that making a similar rescaling could remove the spin effects from $\theta^{\mathrm{Kerr}}$, but in \eqref{Kerrsa} $b$ is already the physical impact parameter and in any case the corresponding diffeomorphism definitely does \emph{not} send the Kerr metric to the Schwarzschild metric!\footnote{Despite this, it is curious to note that the bending of light in Kerr can be derived by a simple replacement $P_{-} \rightarrow M$. See \cite{Frolov:2022rbv} and \cite{Barrabes:2003ey} for more on this point.}


\subsection{The Ferrari-Pendenza metric}

Up to now, we have considered the profile functions which are induced by taking the massless limit of a scattering amplitude which, at leading order, describes the scattering between a probe scalar and Kerr. This reproduced results from the literature for the ultraboost of Kerr which have no spin effects, but they are not the only candidates for the massless limit of Kerr which have appeared in the literature. One potential candidate with a particularly simple profile was obtained long ago by Ferrari and Pendenza~\cite{Ferrari:1990tzs} by ultraboosting along the axis of rotation.

The ultraboosting procedure taken in this paper is somewhat Byzantine, but the resulting line element is in the Kerr-Schild impulsive pp-wave class \eqref{ppwave}, with profile function\footnote{We have normalized the lightfront energy of the solution with a factor of 2 relative to~\cite{Ferrari:1990tzs}.}
\be\label{FP1}
f(x^\perp)=8\,G_{N}P_{-}\,\log(\mu\,|r^2+a^2|)\,,
\ee
where $a:=\sqrt{-a^{\mu}a_{\mu}}$ and $a^{\mu}$ is the usual rescaled Pauli-Lubanski pseudovector. In~\cite{Ferrari:1990tzs}, the stress tensor of this metric is claimed to be
\be\label{FPST0}
T_{--}=P_{-}\,\delta(x^-)\,\delta(r-a)\,,
\ee
corresponding to a null ring of radius $a$. If true, this would certainly be an interesting candidate for the massless limit of Kerr, retaining a finite ring radius with large distance spin effects.

Unfortunately, the claim \eqref{FPST0} seems to be erroneous. A careful calculation of the stress tensor associated with \eqref{FP1}, which we defer to Appendix~\ref{EM-FP}, actually leads to
\be\label{FPST1}
T_{--}=-2P_{-}\,\delta(x^-)\,\frac{a^2}{(r^2+a^2)^2}\,\frac{1}{\pi}\,,
\ee
which is the stress tensor of a null dust, completely de-localized in the transverse plane. As such, this metric is not in our `admissible' class of metrics to describe an ultraboosted black hole, and it is easy to see that it also violates the weak energy condition. More recently, the same line element has been derived using off-shell techniques \cite{Cristofoli}. In this case, the unphysical nature of the solution can be attributed to the ambiguity in the use of off-shell amplitudes, which are not unique. This is why we chose to adopt an on-shell description for shock waves (\ref{eq:eik-pro-map}), in order to avoid this ambiguity.

While it is not appropriate to describe the massless limit of the Kerr metric, it is worth noting that the metric \eqref{FP1} still has some interesting applications. For instance, in celestial holography it corresponds to a conformal primary operator on the celestial sphere~\cite{Pasterski:2020pdk}, and celestial amplitudes computed in this background have surprisingly good convergence properties~\cite{Gonzo:2022tjm} -- probably due to the extended nature of the source \eqref{FPST1}.


\subsection{Ultraboosting the source of Kerr}

At this point, one could be tempted to say that we have exhausted all possibilities for the massless limit of Kerr in the class of metrics \eqref{ppwave}. However, carrying out the ultraboost of any solution at the level of its line element (or metric) is full of ambiguities: depending on the coordinate system various quantities vanish or blow up and choices must be made to obtain an interesting but finite result. Generally, a much less ambiguous option is to ultraboost the \emph{source} of the initial metric, then feed the result back into the Einstein equations to obtain a metric\footnote{It is instructive to work through this in the case of Schwarzschild, where all that is required is a reparametrization of proper time along the worldline of the source in order to obtain the stress tensor \eqref{ASsource} of the shockwave.}.

The source of the Kerr metric was investigated long ago by Israel, who argued that Kerr is sourced by an equatorial disk of mass $M$ whose radius is given by the spin parameter $a$~\cite{Israel:1970kp,Israel:1976vc}\footnote{For related and alternative perspectives on the interpretation of sources in linear and non-linear gravity, see~\cite{Sachs:1958zz,Janis:1965tx,Geroch:1970cd,Perjes:1971gv,Hansen:1974zz,Krasinski:1976vyc,Balasin:1997qk,Burinskii:2001bq}.}. This description was also confirmed in a precise distributional analysis by Balasin and Nachbagauer some years later~\cite{Balasin:1993kf}. Subsequently, these authors computed the ultraboost of this disk-like source parallel to the direction of spin, obtaining the result~\cite{Balasin:1994tb}:
\be\label{BNst}
T_{--}=\frac{P_{-}}{4\,\pi\,a}\,\delta(x^-)\left(\delta(r-a)-\Theta(a-r)\,\frac{r^2}{(a^2-r^2)^{3/2}}\right)\,.
\ee
This is clearly an admissible stress tensor, but -- remarkably -- is \emph{not} the stress tensor of a shockwave, so the resulting metric will certainly differ from what was obtained using the massless limit of the GOV amplitude (or equivalently by na\"ively ultraboosting the metric directly).

Feeding this into the Einstein equations leads to the profile function~\cite{Balasin:1994tb}
\be\label{BNprofile}
f(x^\perp)=8\,G_{N}P_{-}\,\log(\mu\,r)-4G_{N}\,P_{-}\,\Theta(a-r)\left(2\, \log \left(\frac{r}{a+\sqrt{a^{2}-r^{2}}}\right)+\frac{\sqrt{a^{2}-r^{2}}}{a}\right)\,.
\ee
From this, we see that for $r>a$ in the transverse plane the profile is equivalent to that of a non-spinning shockwave. For $r<a$, the $\log(\mu r)$ is removed from the profile function and replaced by some term with a smooth $r\to0$ limit. Remarkably, this profile can also be obtained by carefully ultraboosting the line element of Kerr in $D>4$ (i.e., spin-aligned Myers-Perry) and then taking a smooth $D\to4$ limit~\cite{Yoshino:2004ft}.

\medskip

Therefore, at sufficiently long distances, no spin effects are expected in impact parameter space. In order to set the stage for a discussion of the various scales of the problem in the next section, we now discuss how this scale in impact parameter space translates to a scale in momentum space, via a saddle in the eikonal amplitude. The latter is defined as
\begin{equation}
    \label{eq:eiko-full}
\im \cM_{\mathrm{eik}}(q_{\perp})\sim
2 s 
\int\d^{2}x_{\perp}\,\e^{-\im\,q_{\perp}\cdot x^{\perp} }\,\left(\e^{\im\,\chi_{1}(x_{\perp})}-1\right)\,,
\end{equation}
in terms of the eikonal phase $\chi_1(x^\perp)=-\frac{p_+}{2}f(x^\perp)$, where $s=2p_+ P_-$.

For the spinless case (the standard Aichelburg-Sexl metric), this integral is dominated by a saddle point~\cite{Amati:1987uf,Ciafaloni:2015xsr} located at
\begin{equation}
\label{eq:eikonal-saddle}
|x_\perp|= \frac{
G_N\, s}{|q_{\perp}|}=\frac{G_N\,s}{\sqrt{|t|}}.  
\end{equation}
Since $|t|/s\ll1$, and recalling the $D=4$ Schwarzschild radius $R_S = 2 G_N \sqrt{s}$, this implies that $x^\perp\gg R_S$. Therefore, the integral is safely dominated by a region far away from strong curvature, where the eikonal approximation would break down. This guarantees the consistency of the approximation, and that the eikonal amplitude correctly captures long distance physics. It can then be shown that on the saddle, the eikonal amplitude reduces to
\begin{equation}
  \label{eq:saddle-amp}
  \mathcal{M}_{\mathrm{eik}}(q_{\perp}) \propto  \frac{ s^2}{t}\,\e^{\im \phi}\,,
\end{equation}
which is just the Born approximation dressed by a phase.

For the Balasin-Nachbagauer (BN) profile function \eqref{BNprofile}, if the parameters of the problem are such that the saddle is located at $|x_\perp|>a$, the saddle point is unchanged, and it is clear that the eikonal amplitude, on the saddle, is also unchanged. Using the explicit expression for the saddle \eqref{eq:eikonal-saddle}, $|x_\perp|>a$ implies that for $q$ such that
\begin{equation}
\label{eq:eiko-a}
a |q_\perp| < G_N\, s \,,    
\end{equation}
one should see no spin effects. Conversely, if the saddle is located at $r<a$ ($a |q_\perp| > G_N s$), the incoming particle traverses the BN disk and spin effects are present in the geodesics, and in the eikonal amplitude.

At fixed $s$, $t$, these two regimes can also be written in terms of $a$ compared to the Schwarzschild radius $R_S$:
\begin{eqnarray}
a/R_S  < \sqrt{s/|t|} & \Longrightarrow &\textrm{no spin effects}\nonumber\\
a/R_S  > \sqrt{s/|t|} & \Longrightarrow &\textrm{spin effects}
\end{eqnarray}
One therefore sees that spin effects could be captured at fixed $\sqrt{s/|t|}$ in a regime where $a/R_S$ is large.\footnote{For completeness, note that if $a<R_S$ the finite size effects are hidden within a region which does not contribute to the eikonal physics and cannot be seen in the amplitude.} Whether these effects, or those which we will see in the Born amplitude in the next section, can be related to physical spin effects in a Kerr background is an interesting open question.

%




\medskip

Before turning to this, let us discuss the following issue: how does this construction evade other arguments in the literature which state that the ultraboost of Kerr along the direction of spin is equivalent to a shockwave? In particular, the works~\cite{Griffiths:1997hx,Podolsky:1998rp} analysed the class of stress tensors that can arise from null multipole particles. Using polar coordinates $(r,\phi)$ on the transverse $x^{\perp}$-plane, they argue that a general source with null multipoles takes the form
\be\label{nullmult1}
T_{--}=\delta(x^-)\,\left(-\frac{b_0}{4}\,\delta(r)-\sum_{m=1}^{\infty}\frac{b_m}{4}\,\frac{(-1)^m}{(m-1)!}\,\delta^{(m)}(r)\,\cos\left(m(\phi-\phi_m)\right)\right)\,,
\ee
where the real constants $\{b_{m},\phi_m\}$ describe the $m^{\mathrm{th}}$ multipole of the null source, and $\delta^{(m)}$ denotes the $m^{\mathrm{th}}$-derivative of the delta function.

With this general ansatz, it is straightforward to show that when ultraboosting the Kerr metric along its axis of rotation, one obtains~\cite{Podolsky:1998rp}
\be\label{nullmult2}
b_{0}=-4\,P_{-}\,, \qquad b_{m}=0=\phi_{m} \quad \forall m>0\,,
\ee
which immediately reduces the source to that of the shockwave. However, it is not the case that every function describing a compactly supported source can be decomposed in terms of the Dirac delta function and its derivatives. Indeed, the ansatz \eqref{nullmult1} implies that the source is localized on formal neighborhoods of a point in the transverse plane, restricting the structure of the source to have only `infinitesimal' finite size.

By contrast, the Heaviside theta function appearing in the second term of \eqref{BNst} cannot be decomposed in terms of delta functions and their derivatives, and encodes an extended object of finite size in the transverse plane. Thus, by ultraboosting the (distributional) source of Kerr parallel to the direction of spin, a source can be obtained which avoids the restrictive multipolar structure \eqref{nullmult2} by simply lying in a more general function space (cf., \cite{Aoude:2020ygw,Guevara:2020xjx,Aoude:2021oqj}).


\section{Scattering off the ultraboosted source of Kerr}
\label{Sec:BN}

Of all the candidates for the massless limit of Kerr considered in the previous section, the BN metric \eqref{BNprofile}, obtained by ultraboosting the source, is clearly the most interesting. This is an admissible metric with compactly supported source and no long-range spin effects, but differs from the shockwave at small distances with explicit dependence on the spin parameter. Scattering a scalar probe off of this metric at generic impact parameter should therefore produce an eikonal phase, and hence 4-point scattering amplitude, with explicit spin effects. 

In this section, we analyse the Born amplitude $A_4$ associated with this metric, the physical scales in play between the Born and eikonal amplitudes, and the UV-properties of the Born amplitude, all of which have interesting, non-trivial features.


\subsection{Born amplitude}
\label{sec:Born-Amp}

Scattering a scalar probe on any metric of the form \eqref{ppwave} will always have an eikonal-like structure~\cite{tHooft:1987vrq,Adamo:2021rfq}, controlled by a phase $\chi_1$. While the interpretation of this phase as a true eikonal phase is subtle in general, one can proceed na\"ively an Fourier transform to obtain an associated. four-point Born amplitude.


With this in mind, the four-point amplitude associated to the BN metric is
\be\label{eq:four-poin-BN}
A_{4}(q_{\perp})=-2p^2_{+}P_{-}\, \int \mathrm{d}^{2} x^{\perp} \,  \e^{-\im\, q_{\perp} x^{\perp}}\, f(x^{\perp}) \, ,
\end{equation}
where the profile is given by \eqref{BNprofile}. The result can be split into a spinless and spinning contribution:
\begin{equation}\label{BNamp1}
A_{4}(q_{\perp})=\frac{32 \pi G_{N}\, p^2_{+}\,P_{-}^2}{q^2_{\perp}}+16\pi\,G_{N} p^2_{+}\,P_{-}^2 \: \mathcal{I}(q_{\perp},a) \, ,
\end{equation}
where we have defined
\begin{equation}
\mathcal{I}(q_{\perp},a) := \int_{0}^{a} \d r\,r\, J_{0}(|q_{\perp}|\,r)\left(2 \log \left(\frac{r}{a+\sqrt{a^{2}-r^{2}}}\right)+\frac{\sqrt{a^{2}-r^{2}}}{a}\right) \, ,
\end{equation}
with $J_0$ a Bessel function of the first kind. The remaining integral can be performed by first changing variables to $x=r/a$, giving
\begin{equation}
\mathcal{I}(q_{\perp},a) =\frac{\sin(a|q_{\perp}|)-a|q_{\perp}|\,\cos(a|q_{\perp}|)}{a\, |q_{\perp}|^3}+2a^2 \int_{0}^{1} \d x\,  x \, J_{0}(|q_{\perp}|ax)\, \log \left(\frac{x}{1+\sqrt{1-x^{2}}}\right) \, .
\end{equation}
Splitting the logarithm into two terms leaves 
\begin{multline}\label{eq:int-in-BN}
\mathcal{I}(q_{\perp},a) =\frac{\sin(a|q_{\perp}|)-a|q_{\perp}|\,\cos(a|q_{\perp}|)}{a\, |q_{\perp}|^3}+\frac{2}{q^2_{\perp}}\,\left(J_{0}(a|q_{\perp}|)-1\right)  \\
-a^2 \int_{0}^{1} \d y\, J_{0}(|q_{\perp}|\,a\sqrt{y}) \, \log \left(1+\sqrt{1-y}\right) \,,
\end{multline}
where the remaining integral can now be evaluated using the definition of the Bessel function.

This gives us the following compact result:
\begin{equation}\label{eq:Born-BN}
\mathcal{I}(q_{\perp},a) =\frac{\sin(a|q_{\perp}|)-a|q_{\perp}|\,\cos(a|q_{\perp}|)}{a\, |q_{\perp}|^3}-\frac{4\sin^2(\frac{|q_{\perp}|a}{2})}{q^2_{\perp}}  \, ,
\end{equation}
where the Bessel function in the first line of \eqref{eq:int-in-BN} has been cancelled by a contribution equal and opposite coming from the remaining integral. After combining this result with the non-spinning term in \eqref{BNamp1} and applying some basic trigonometric identities, we are left with a remarkably simple formula for the four-point amplitude:
\begin{equation}\label{BNamp}
A_{4}(q_{\perp})=\frac{\kappa^2\,(p_+\,P_-)^2}{2\,q^2_{\perp}}\, \left(\frac{\sin(a |q_{\perp}|)}{a\, |q_{\perp}|}+\cos(a |q_{\perp}|) \right) \,,
\end{equation}
where we have reinstated $\kappa^2=32\pi G_{N}$.

In the $a\to0$ limit, this amplitude reduces to the gravitational tree amplitude between scalars. At generic $a,q_\perp$, the amplitude $s^2/t$ receives infinitely many corrections of the form $a^{2n}t^n$. It remains a tree-level (meromorphic) object, but the higher order terms describe finite size effects associated to the multipoles of the BN metric. This is further clarified by examining the stress tensor \eqref{BNst} associated with the solution. In particular, this corresponds to a disc of null dust bounded by a ring singularity of radius $r=a$ in the $x^-=0$ null plane, with a density profile inside of the disc given by the factor multiplying the step function in \eqref{BNst}. This is a null version of Israel's distributional source for the Kerr metric itself~\cite{Israel:1970kp}. It is remarkable that the scattering amplitude between a scalar and such a finite-size object takes a form as simple as \eqref{BNamp}. 



\medskip

An immediate worry that one could have when looking at \eqref{BNamp} is that amplitude might violate polynomial boundedness in $s$ (associated with causality) after applying crossing symmetry. However, the eikonal limit is a resummation that describes small $t$ and should not satisfy crossing symmetry, therefore the absence of polynomial boundedness is not expected. In particular, even the standard, non-spinning eikonal amplitude is not polynomially bounded~\cite{tHooft:1987vrq}. Incidentally, note that the spinning amplitude \eqref{eq:GOV-v2} with finite mass already has an exponential in $q$, so would have the same acausal behaviour under crossing symmetry. Here too, crossing symmetry is not expected to hold as-is, and a UV completion of the amplitude -- if it exists and can be made crossing symmetric -- should provide terms which remedy this behaviour. 

Note also that this amplitude does not have sufficiently many terms in its Laurent expansion to be checked against positivity constraints (cf., ~\cite{Bellazzini:2020cot,Tolley:2020gtv,Caron-Huot:2020cmc,Herrero-Valea:2020wxz,Arkani-Hamed:2020blm,Chowdhury:2021ynh,Bern:2021ppb}), though it would be interesting to further explore this using `reverse bootstrap' ideas~\cite{Alberte:2021dnj} or the non-projective EFT-hedron~\cite{Chiang:2022ltp}. Since it is fair to expect that our amplitude will show signs of acausal behaviour (due to its relation to over-spun Kerr), it could suggest derivations for further positivity constraints which could be used to discard the amplitude on causality grounds.


\subsection{Physical scales: Born vs. eikonal}

For the Born amplitude \eqref{BNamp}, we see that there are no spin effects only when $a$ is such that $a |q_{\perp}|\ll1$. In particular, spin effects turn on for $a |q_{\perp}|>1$. It is interesting to compare this with the analysis of the eikonal saddle above. Given that $G_Ns\gg1$, we have the situation described in Figure~\ref{fig:scales}:
\begin{figure}[h]
    \centering
    \includegraphics{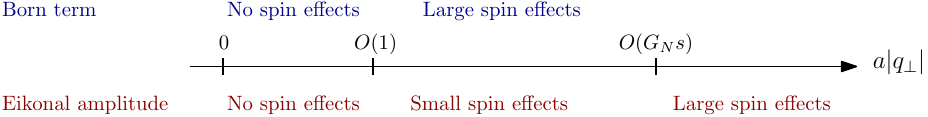}
    \caption{Scales in $a|q_{\perp}|$ at which spin effects occur for the Born and eikonal amplitudes of the BN metric.}
    \label{fig:scales}
\end{figure}

This picture seems quite counter-intuitive: ordinarily, as one passes from very large to small impact parameters, the amplitude that dominates first is the Born term. Then, at smaller distances, the eikonal phase becomes of order $1$ and the Born term gets dressed by the eikonal phase~\cite{Amati:1987uf,Giddings:2007qq,Giddings:2009gj,Giddings:2011xs}. This can be made fully precise in dimension\footnote{In higher dimensions the BN profile differs from what it is in $D=4$~\cite{Yoshino:2004ft}, so the discussion above relied on the assumption that the equation \eqref{eq:Born-BN} can be interpreted as a higher-dimensional amplitude.} $D>4$ where the eikonal phase is given by (up to normalisation terms)
$\chi_1 = G_N \frac{s}{b^{D-4}}$
and it becomes of order one when 
$    b\sim(G_N s)^{1/(D-4)}=(E R_S^{D-3})^{1/(D-4)}$
where $E$ is the centre of mass energy and the Schwarzschild radius in $D$ dimensions is 
$    R_S^{D-3} = 2 G_N E\,.$
In $D=4$ the phase is infinite due to an IR divergence (regulated by the mass scale $\mu$), but even after subtraction of the infinite constant it is never small and is given by $s \log(t)$.

In this context, Figure~\ref{fig:scales} is puzzling because at larger distances (Born distances) one sees spin effects, while at eikonal distances, one sees fewer spin effects.

The resolution of this conundrum is as follows. In $D>4$, in the regime of impact parameters where the Born term dominates, the relevant regime of $q_\perp$ is such that $a q_\perp\ll1$, so that the Born term physics does not produce spin effects. Then, as the impact parameter decreases, the relevant range of $q_\perp$ increases, so as to eventually allow for spin effects. In $D=4$, the eikonal phase is never small, so presumably the Born term is never a good physical approximation of the scattering amplitude.

Of course, one could question the validity of the saddle-point approximation that determines the eikonal amplitude. In appendix~\ref{sec:toy-model}, we provide an example of a toy-model for the BN metric for which the eikonal amplitude can be computed exactly. This confirms the reasoning of this section: namely, that no spin corrections are expected before $a |q_\perp| \sim G_N s$.


\subsection{UV properties}

Notwithstanding these observations, the Born amplitude associated with the BN metric \eqref{BNamp} is still an interesting object in its own right. From the start we knew that something had to be un-physical about the limits which produced the metric, as the Kerr black hole is over-spun in the process. However, this large spin limit is also taken for instance in~\cite{Arkani-Hamed:2019ymq} and might still be the correct way to capture a certain class of spin corrections to the shockwave. With this in mind, we might as well proceed to study the amplitude on its own and see what properties it has. 

One obvious property of the Born amplitude \eqref{BNamp} is the presence of oscillations due to the trigonometric functions, similar to that of a phase. On par with the unitarisation produced by the eikonal phase at high energies~\cite{Giddings:2011xs}, one might expect that this amplitude has better UV behaviour than the non-spinning one. Doing the same approximation as above (i.e., assuming that the amplitude \eqref{BNamp} is valid in higher dimensions), we investigated the partial wave expansion of the amplitude. Without spin, for the $s^2/t$ amplitude, the partial wave coefficients $a_J(s)$ behave as $s^{D/2-1}$ (independently of $J$). Adding spin softens this to $a_J(s)\sim s^{D/4-1/2}$, as can be seen by computing these coefficients explicitly; more details on the partial wave expansion are given in appendix~\ref{sec:pw}.

An interesting consequence of this UV softening is that the Born amplitude can be Mellin transformed to the celestial sphere. One finds the corresponding celestial amplitude
\begin{equation}\label{BNcelestial}
    \int _0^{\infty }\omega ^{\beta -1}\,A_4\!\left(\omega ^2,-z\omega ^2\right)\,\d\omega = \frac\beta2\,  a^{-\beta -2}\, \cos(\pi\beta/2)\, \Gamma (\beta +1)\, z^{-\frac{\beta }{2}-2}\,.
\end{equation}
This amplitude has poles at $\beta=-2n$ for $n$ a positive integer, reflecting IR physics~\cite{Arkani-Hamed:2020gyp}, with residues proportional to $a^{2n-2} z^{n-2}$. Further, the amplitude is regular in the right-side of the complex $\beta$ plane, as expected from QFT considerations.

By contrast, the tree-level graviton exchange amplitude between scalars cannot be transformed to the celestial sphere as its UV and IR behaviours make the Mellin transform divergent. For the BN amplitude \eqref{BNcelestial}, the UV is regulated and the transform can converge in some range of parameters (away from the poles). A similar effect was also observed by~\cite{Gonzo:2022tjm} for probe scattering in the Ferrari-Pendenza metric, where the source has \emph{infinite} size in the transverse plane.

Note that the Mellin transform of the toy-model amplitude used in appendix~\ref{sec:toy-model} can also be easily Mellin transformed, giving rise to the amplitude
\begin{equation}
2^{\beta +1}\, a^{-\beta -2}\,z^{-\frac{\beta }{2}-2}\,    \frac{ \Gamma\! \left(1+\frac{\beta }{2}\right) }{\Gamma\! \left(-\frac{\beta }{2}\right)}\,,
\end{equation}
which has the same characteristics as its BN cousin above; in particular, it has poles at $\beta=-2n$ with residues proportional to $a^{2n-2} z^{n-2}$. This reinforces the conjecture that these amplitudes exist because of the generic feature that their UV behaviour is softened by the extended nature of the objects involved in the scattering.


\section{Single copy of the ultrarelativistic limit of Kerr}
\label{Sec:SingleCopy}

Our amplitudes-based approach to the massless limit of Kerr suggests that there should be some double copy structure in play\footnote{Of course, it is not the tree-level scattering amplitude but the eikonal phase which directly determines the metric profile, and details of the double copy in the eikonal regime are quite subtle (cf., \cite{Saotome:2012vy,Naculich:2020clm,Naculich:2022npm}).}. For finite mass, the framework of classical double copy~\cite{Monteiro:2014cda,Monteiro:2020plf} indicates that the `single copy' of the Kerr metric in electromagnetism is a spinning point charge, often called the $\sqrt{\mathrm{Kerr}}$ solution~\cite{Newman:1965tw,Monteiro:2014cda}. Thus, one can naturally ask: what is the ultrarelativistic limit of $\sqrt{\mathrm{Kerr}}$, and is it the single copy of the ultrarelativistic limit of Kerr?

To begin, we can take a purely amplitudes-based approach to this question by applying an electromagnetic version of our fundamental relation \eqref{eq:eik-pro-map}. Firstly, we will consider electromagnetic fields in Minkowski spacetime, where the gauge potential can be put into the form
\be\label{EM1}
A=\delta(x^-)\,\varphi(x^{\perp})\,\d x^{-}\,,
\ee
in lightfront coordinates; the function $\varphi(x^\perp)$ is the profile of the electromagnetic field. As in the gravitational case, such fields have a single source component:
\be\label{EMsource}
J_{-}=\delta(x^-)\,\partial_{\perp}^2\varphi(x^\perp)\,,
\ee
and we will call the field admissible if this source is compactly supported.

Now, the electromagnetic version of the relation \eqref{eq:eik-pro-map} is easily shown to be
\be\label{EMeik}
\varphi(x^\perp)=-\frac{2\,\chi_1(x^\perp)}{e}\,,
\ee
where $e$ is the charge of a probe scalar and $\chi_1$ is the eikonal phase associated with electromagnetic scattering between the probe and the source of the gauge field. Proceeding as we did in the gravitational case, the tree-level, four-point scattering amplitude between a charged massive scalar and a massive charged particle with infinite spin in electromagnetism is~\cite{Arkani-Hamed:2019ymq}:
\be\label{EMGOV1}
A_{4}(q)=-2\,e\,Q\,\frac{p\cdot P}{q^2}\,\sum_{\pm}(1\pm v)\,\exp\left(\pm\im \frac{q\cdot[w*s]}{M}\right)\,,
\ee
where $Q$ is the charge of the spinning particle.

Taking the massless limit of the associated eikonal phase (using the same scaling rules as in the gravitational case) leads to
\be\label{EMGOV2}
\lim_{M\to0}\chi_{1}(b^{\perp})=-\frac{e\,Q}{4\,\pi}\,\log\left(\mu\,\sqrt{(b-[w*a])_{\perp}^2}\right)\,.
\ee
Feeding this into \eqref{EMeik} leads to profiles
\be\label{EMGOV3}
\varphi(x^\perp)=\frac{Q}{2\,\pi}\,\log\left(\mu\sqrt{(x-\tilde{a})_{\perp}^2}\right)\,,
\ee
from which it is obvious (by the same reasoning as the gravitational case) that there will be no spin effects regardless of the direction of the ultraboost. That is, in each case the profile is equivalent to that of an electromagnetic shockwave.

\medskip

As the Balasin-Nachbagauer solution \eqref{BNprofile} seems to be the most interesting candidate for the massless limit of Kerr, it is natural to consider the single copy version of this solution in electromagnetism. By simply applying the Kerr-Schild classical double copy map~\cite{Monteiro:2014cda} to \eqref{BNprofile}, one is left with a gauge potential of the form \eqref{EM1} with profile
\be\label{KSDC}
\varphi(x^\perp)=\frac{Q}{2\pi}\left[ \log(\mu r)-\Theta(a-r)\left(\log \left(\frac{r}{a+\sqrt{a^{2}-r^{2}}}\right)+\frac{\sqrt{a^{2}-r^{2}}}{2a}\right)\right] \, .
\ee
As expected, at distances $r>a$ this corresponds to an electromagnetic shockwave, while for $r<a$ the solution describes a disc of radius $a$ with total charge $Q$~\cite{Newman:1965tw}.

It is worth noting that, at least na\"ively, the direct ultraboost of the source of $\sqrt{\mbox{Kerr}}$ obtained by amplitudes methods does \emph{not} produce the same profile with finite size effects. Indeed, by using a Fourier transform seeded with the $1\to2$ amplitude for a spinning point charge emitting a photon (written in split signature or with complex kinematics to ensure a non-vanishing result), the source for $\sqrt{\mbox{Kerr}}$ can be written as~\cite{Vines:2017hyw,Monteiro:2021ztt}
\be\label{RKSource}
J_{\mu}=Q\int\d\tau\,U^{\nu}\,\exp(a*\partial)_{\mu\nu}\,\delta^{4}(x-U\,\tau)\,,
\ee
where $U^{\mu}$ is the 4-velocity of the source and 
\begin{equation}
 (a*\partial)_{\mu\nu}\equiv\epsilon_{\mu\nu\rho\sigma}\,a^{\rho}\,\partial^{\sigma}\,.   
\end{equation}
It is straightforward to ultraboost this solution with $U^{\mu}\to n^{\mu}$ and $a^{\mu}$ fixed, resulting in a source current with a single lightfront component
\be\label{RKbS1}
J_{-}=Q\,\delta(x^-)\,\exp(a_\perp\wedge\partial_\perp)\,\delta^{2}(x^\perp)\,,
\ee
with $a_{\perp}\wedge\partial_\perp:=a_{1}\partial_2-a_2 \partial_1$ acting in the transverse plane.

This source actually lies in the electromagnetic version of the `null multipole' class \eqref{nullmult1}, since $J_-$ can be expanded as
\be\label{RKbS2}
J_{-}=Q\,\delta(x^-)\left[\delta(r)+\sum_{m=1}^{\infty}\frac{\left(a_{\perp}\wedge\hat{x}_{\perp}\right)^{m}}{m!}\,\delta^{(m)}(r)\right]\,, \quad \hat{x}_{\perp}:=\frac{1}{r}\,(x_{1},\,x_{2})\,.
\ee
As such, it describes an electromagnetic source localized in formal neighbourhoods of the point $r=0$ in the transverse plane: each term in the sum represents higher and higher infinitesimal deformations (by $a_{\perp}$) away from this point. However, this tower of infinitesimal deformations does not lead to the extended, finite size effects produced by the theta function appearing in \eqref{KSDC}.


\section{Discussion}

In this paper, we have investigated the ultrarelativistic limit of the Kerr metric within the class of Kerr-Schild impulsive pp-waves, using an on-shell relationship between the metric profile and the eikonal phase associated with scattering between a probe scalar and the source of the metric. In all cases arising from scattering amplitudes or the literature, we found that there are no true spin effects at sufficiently large distances after taking the ultrarelativistic limit. However, one metric -- obtained by Balasin and Nachbagauer (BN) by ultraboosting the \emph{source} of Kerr rather than the line element itself -- has interesting finite size effects at sufficiently small distances, sourced by a disc of null dust with finite radius and energy.

The BN metric leads to a remarkably compact tree-level Born amplitude \eqref{BNamp} which exhibits finite size spin effects at sufficiently small impact parameters. We analysed the behaviour of this amplitude, which improved high-energy behaviour due to softening by oscillations induced by the finite size effects. This led to an apparent conundrum when comparing the on-set of spin effects in the eikonal regime, from which we concluded that this Born amplitude is not actually a good physical approximation to the scattering amplitude. Intuitively, this is perhaps not totally surprising: the Born amplitude is obtained by a Fourier transform of an eikonal phase which is equal (up to normalization) to the BN metric profile. However, the relation between the eikonal phase and metric profile is really only valid in a regime where the source appears point-like (i.e., the leading eikonal approximation), so the Fourier transform should be cut-off to restrict to this regime.

Nevertheless, it would be interesting to further explore the extent to which the tree-level BN amplitude is able to capture relevant spin effects in the ultrarelativistic limit. That is, just because the Born approximation is not totally physical doesn't mean that it cannot encode some of the classical spin effects associated with Kerr (cf., ~\cite{Moynihan:2019bor,Bautista:2021wfy,Haddad:2021znf,Chen:2021kxt,Aoude:2022trd,Aoude:2022thd,Menezes:2022tcs,Damgaard:2022jem}). By analogy, we saw that the ultraboost of Kerr orthogonal to the axis of rotation did not contain true spin effects, but a naive replacement rule nevertheless correctly reproduces the bending of light in the Kerr metric~\cite{Barrabes:2003ey,Frolov:2022rbv}. 

\medskip

There is a natural generalization beyond the class of metrics we considered which could better capture the ultrarelativistic limit of the Kerr metric. The set of `impulsive gyraton' metrics inside of the pp-wave class has three (rather than one) functional degrees of freedom in $D=4$, and these certainly encode internal angular momentum~\cite{Bonnor:1970sb,Frolov:2005zq,Frolov:2005in,Podolsky:2014lpa,Frolov:2022rbv}. Thus, it could be that a richer candidate for the ultrarelativistic limit of Kerr can be found in this impulsive gyraton class, although the straightforward relationship between the metric and on-shell scattering amplitudes that was utilized here would need to be modified. 

\medskip

There are also several interesting open questions associated with the BN Born amplitude, viewed simply as a formal scattering amplitude. A first (somewhat technical) step would be to attempt a calculation of the BN eikonal amplitude. We have not yet found a straightforward way of doing this, but it could be that a holomorphic factorization argument, akin to what is required to evaluate the eikonal amplitude associated with the Kerr metric~\cite{Adamo:2021rfq}.

However, since the UV behaviour of the tree-level BN amplitude is essentially regularized by finite size effects, it is tempting to ask if this is actually more than just a Born amplitude. That is, can \eqref{BNamp} be interpreted as some sort of (quasi-)eikonal amplitude in its own right? This could provide a way to ascribe a truly physical interpretation to the amplitude, which seems to be precluded if we interpret it in the strict Born approximation.

Conversely, one could try to rule out any physical interpretation of the BN amplitude, for instance using causality and unitarity constraints. As it stands, \eqref{BNamp} has non-polynomial behaviour under $t$-channel crossing in the complex plane, which would lead to violent causality violation. As discussed in Section~\ref{Sec:BN}, this is not a problem in the eikonal regime (where polynomial behaviour under crossing is not expected), but if a causal crossing-symmetric unitary UV-completion exists for this amplitude, it must -- after crossing -- regulate the exponential divergence coming from the cosine and sine functions. One could also try to crossing-symmetrize the amplitude by hand, and see if the low energy expansion coefficients satisfy positivity constraints~\cite{Bellazzini:2020cot,Tolley:2020gtv,Caron-Huot:2020cmc,Herrero-Valea:2020wxz,Arkani-Hamed:2020blm,Chowdhury:2021ynh,Bern:2021ppb}. However, it turns out that the resulting amplitude does not possess enough of these low energy coefficients to draw a conclusion on this point.

Finally, it would be interesting to study the emission of gravitational radiation involving the BN metric. A first approximation to this would be to consider the graviton emission amplitude for a probe scalar in the background of a BN spacetime, which corresponds to wave emission from the probe itself. A more in-depth computation for the ultrarelativistic regime would involve computing the metric describing a collision between the BN metric and a scalar shockwave, and computing the resulting news tensor at future null infinity (akin to what has been done for the collision of two shockwaves~\cite{Gruzinov:2014moa}).


\acknowledgments We thank Matteo Sergola, Justin Vines and Sasha Zhiboedov for helpful discussions and Matteo Sergola for comments on a draft. TA is supported by a Royal Society University Research Fellowship and by the Leverhulme Trust (RPG-2020-386). AC is supported by the Leverhulme Trust (RPG-2020-386).

\appendix

\section{The stress-energy tensor of the Ferrari-Pendenza metric}
\label{EM-FP}

For an ultraboost of Kerr along the axis of rotation, Ferrari and Pendenza claimed a line element~\cite{Ferrari:1990tzs}
\begin{equation}\label{FP-gyra}
\d s^{2}=\eta_{\mu\nu}\,\d x^{\mu}\,\d x^{\nu}+8\, G_{N}\,P\, \delta(n \cdot x)\, \log \left( \mid x^{2}+a^{2} \mid\right) \left(n_{\mu}\, \d x^{\mu}\right)^2 \ ,
\end{equation}
where $n^{\mu}$ is the null vector corresponding to the direction of the ultraboost, $P$ is the finite (lightfront) energy of the solution and $a:=\sqrt{-a^{\mu}a_{\mu}}$ and $a^{\mu}$ is the usual rescaled Pauli-Lubanski pseudovector (as we are working in mostly-minus signature). Here, we have kept the notation generic, to more closely mimic what appears in the original paper, but it is straightforward to convert between this an the explicit lightfront coordinates used in the main text.

Our goal is to compute the stress-energy tensor of this metric. As the line element \eqref{FP-gyra} is Kerr-Schild, introduce
\begin{equation}
h_{\mu \nu}=8\, G_{N}\,P\, \delta(n \cdot x)\, \log \left( \mid x^{2}+a^{2} \mid\right)\, n_{\mu} n_{\nu}\,, 
\end{equation}
from which the stress-energy tensor is identified by feeding \eqref{FP-gyra} into the Einstein equations:
\begin{equation}
T_{\mu \nu}=-\frac{1}{16 \pi G_{N}} \: \Box h_{\mu \nu} \,,
\end{equation}
where $\Box:=\eta^{\mu\nu}\partial_{\mu}\partial_{\nu}$ is the flat spacetime wave operator. The computation is facilitated by making the decomposition
\begin{equation}
h_{\mu \nu}=8\, G_{N}\,P\, n_{\mu}\, n_{\nu}\, \Phi(x)\,,  \qquad  \Phi(x):= \delta(n \cdot x)\, \log \left( \mid x^{2}+a^{2} \mid\right) \, ,
\end{equation}
to isolate the scalar $\Phi(x)$ on which the wave operator acts non-trivially.

Now, it follows that
\begin{equation}
\begin{split}
\Box \Phi(x)&=\int \frac{\d t}{2\pi} \: \Box \: \bigg[\e^{\im\, t\, n \cdot x}\,  \log \left( \mid x^{2}+a^{2} \mid\right)\bigg] \\
&=4\,\delta(n \cdot x) \left[\frac{1}{x^2+a^2}+\frac{a^2}{(x^2+a^2)^2} \right]+4\im \int \frac{\d t}{2\pi} \:  \: \e^{\im\, t\, n \cdot x}\,n\cdot x \, \frac{t}{x^2+a^2} \\
&=4\,\delta(n \cdot x) \left[\frac{1}{x^2+a^2}+\frac{a^2}{(x^2+a^2)^2} \right]+ n \cdot x\, \frac{\d \delta(n \cdot x)}{\d (n \cdot x)}\,\frac{4}{x^2+a^2} \, .
\end{split}
\end{equation}
Using the distributional identity
\begin{equation}
\delta(n \cdot x)+n \cdot x\, \frac{\d \delta(n \cdot x)}{\d (n \cdot x)}=0 \, ,
\end{equation}
we immediately arrive at
\begin{equation}
\Box \Phi(x)=\delta(n \cdot x) \frac{4\,a^2}{x^2+a^2} \, ,
\end{equation}
from which we obtain the stress-energy tensor: 
\begin{equation}\label{eq:t-mu-nu-FP}
T_{\mu \nu}=-2\,P\,\delta(n \cdot x)\, \frac{n_{\mu}\,n_{\nu}\, a^2}{(x^2+a^2)^2}\, \frac{1}{\pi} \ .
\end{equation}
This result disagrees with \eqref{FPST0}, which was claimed in the original paper~\cite{Ferrari:1990tzs}.

In particular, the result \eqref{eq:t-mu-nu-FP} is delocalized in the transverse plane orthogonal to $n^{\mu}$, and it also violates the weak energy condition: for a time-like observer with worldline $x^{\mu}(\sigma)$ and 4-velocity normalized to $u^{\mu}=(1,0,0,0)$, it follows that
\begin{equation}
T_{\mu \nu}(x(\sigma))\,u^{\mu}\,u^{\nu}<0\,, \qquad \forall \sigma \in \mathbb{R}\, .
\end{equation}
This demonstrates that the line element first presented in~\cite{Ferrari:1990tzs} is not in our admissible class of metrics, as it describes a delocalized null dust with negative energy.


\section{Toy model eikonal caluclation}
\label{sec:toy-model}

\begin{figure}
    \centering
    \includegraphics{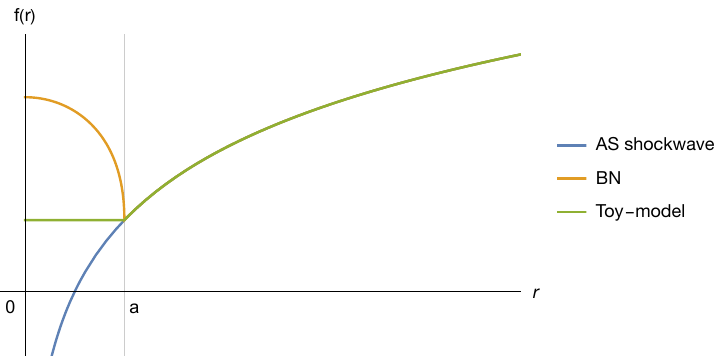}
    \caption{Three different metric profiles}
    \label{fig:profiles}
\end{figure}

As a toy model for the eikonal BN amplitude, we look at the eikonal amplitude corresponding to the following metric profile:
\be\label{Toyprofile}
f_{\mathrm{TM}}(x^\perp)=8\,G_{N}\,P_{-}\,\left[\log(\mu\,r)-\Theta(a-r)\, \log \left(\frac{r}{a}\right) \right]\,.
\ee
This mimics key qualitative properties of the BN metric, being equivalent to the shockwave for $r>a$ with finite size effects (controlled by a Heaviside theta function) removing the logarithm for short distances $r<a$, but has a simple constant profile inside of the ring radius which renders the calculation of the associated eikonal amplitude tractable. See Figure~\ref{fig:profiles} for a comparison between the profiles of the shockwave, BN metric and this toy model.

\def\as{\alpha(s)}
The Born amplitude of the toy model profile is given by
\begin{equation}
    A_{\mathrm{TM}}(q) = -\frac{s^2}{q_{\perp}^2}\,J_0(a|q_{\perp}|)
\end{equation}
where $J_0$ is a Bessel function of the first kind. This is has a structure very similar to that of eq.~\ref{eq:Born-BN}, with the standard $1/t=1/|q_\perp|^2$ pole and an oscillating factor. A simple calculation gives the corresponding exponentiated eikonal amplitude
\begin{equation}
    \im \mathcal{M}_{\textrm{eik}} =
    \frac{2^{1+\im \as}\, \Gamma \left(\frac{\im \as}{2}+1\right) q^{-2-\im \as}}{\Gamma \left(-\frac{\im \as}{2}\right)} + 
    \frac{\as\, a^{2+\im \as}\,  {}_1F_2\left(\frac{\im \as}{2}+1;\,2,\,\frac{\im \as}{2}+2;\,-\frac{a^2}{4} q^2\right)}{2 (\as-2 \im)}\,,
\end{equation}
where $\alpha(s):=G_N\,s$ and ${}_1F_2$ denotes the generalized hypergeometric function.

\begin{figure}
    \centering
    \includegraphics{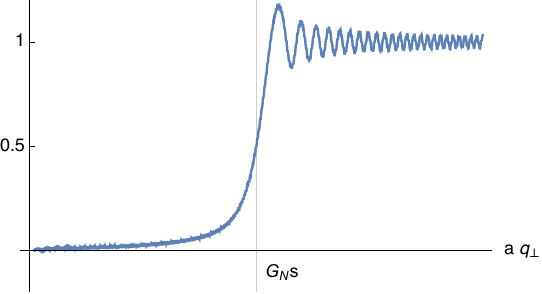}
    \caption{Plot of the ratio of the spin correction to the leading order term. Those grow to reach order 1 correction around $a q_\perp \sim G_N s$. To produce this specific plot, we used typical values of $s$ and $a$, $G_N=1,\,s=500,\,a=10$.}
    \label{fig:my_label}
\end{figure}

Observe that spin/finite size effects are present only in the second term of this eikonal amplitude. One can then consider the ratio of the second and first terms to test the onset of spin effects as $a|q_\perp|$ varies; see Figure~\ref{fig:my_label}. In agreement with the saddle point analysis of the BN metric, we see that spin effects due to the second term only lead to corrections to the first term when $a|q_\perp|\sim G_{N} s$.

\section{Partial wave decomposition}
\label{sec:pw}

Following the conventions of~\cite{Soldate:1986mk}, for the scattering of massless particles at physical $s>0$ and $t=-s \sin^2(\theta)$, where $\theta$ is the scattering angle, we have
\begin{equation}
    A_{4}(s,t) =\lambda_D\, s^{D/2-2}\, \sum_{J=0}^\infty\frac{C_J^\nu(1)}{N_J^\nu}\,  a_J(s)\, C_J^\nu (\cos(\theta))\,,
\end{equation}
where $D$ is the number of spacetime dimensions, $a_J(s)$ are the partial wave coefficients, $\nu:= (D-3)/2$, and $C_J^{\nu}$ are the Gegenbauer polynomials. The overall constant $\lambda_D$ is given by
\begin{eqnarray}
\lambda_D &=& 2\,\Gamma\!\left(\frac{D-1}{2}\right)\, (16\pi)^{D/2-1}\,,
\end{eqnarray}
and the normalisation $N_J^\nu$ is defined through the orthogonality relation 
\begin{equation}
    \int_{-1}^1 \d x\, (1-x^2)^{D/2-2}\,C_J^\nu(x)\, C_{J'}^{\nu'}(x) = N_J^\nu\, \delta_{J,J'}\,\delta_{\nu,\nu'} = \frac{2^{1-2\nu}\,\pi\, \Gamma(J+2\nu)}{J!\, (\nu+J)\,\Gamma^2(\nu)}\,,
\end{equation}
for the Gegenbauer polynomials.

Using the orthogonality relation, one can obtain the partial wave coefficients from integrals of the amplitude:
\begin{equation}
a_J(s) = \frac{2^{D-3}\,s^{D/2-3}}{\lambda_D\, C_J^\nu(1)}\int_{-s}^0\d t\,\left(-\frac{t}{s}-\frac{t^2}{s^2}\right)^{D/2-2}\, C_J^\nu(1+2t/s)\, A_{4}(s,t)\,,
\end{equation}
where the normalisation constant $N_J^\nu$ are dropped. Note that this definition for the coefficients $a_J(s)$ makes them dimensionless, and unitarity of the S-matrix partial waves would be expressed in terms of $S_J(s) = 1+2\im\, a_J(s)$ as $|S_J(s)|^2<1$. For more details, see the review \cite{Correia:2020xtr}, where the reader should pay attention to the difference of normalisation. 

These definitions give rise to the results mentioned in the text. Note that these normalizations differ from those of \cite{Correia:2020xtr} for instance, where the factor of $s^{D/2-1}$ is  included in the definition of the partial wave coefficients, which are called $f_J$ there.

\bibliography{NewBib}
\bibliographystyle{JHEP}

\end{document}